\documentclass[12pt]{iopart}

\usepackage{graphicx}
\usepackage{epsfig}
\usepackage{iopams}
\newcommand{\sgn}{{\rm sgn}\,}
\newcommand{\im}{{\rm Im}\,}

\newcommand{\Str}{{\rm Str}\,}
\newcommand{\Sdet}{{\rm Sdet}\,}
\newcommand{\diag}{{\rm diag}\,}

\begin{document}

\title{Exact Coupling Coefficient Distribution in the Doorway Mechanism}

\author{Heiner Kohler$^{(1)}$, Thomas Guhr$^{(1)}$ and 
Sven \AA berg$^{(2)}$}

\address{
$^{(1)}$Fachbereich Physik, Universit\"at Duisburg--Essen, Duisburg, 
Germany \\
$^{(2)}$Matematisk Fysik, Lunds Universitet, Lund, Sweden}
\ead{heinerich.kohler@uni-due.de}

\begin{abstract}
In many--body and other systems, the physics situation often allows one
to interpret certain, distinct states by means of a simple picture. In this
interpretation, the distinct states are not eigenstates of the
full Hamiltonian. Hence, there is an interaction which makes the distinct
states act as doorways into background states which are modeled 
statistically. The crucial
quantities are the overlaps between the eigenstates of the full Hamiltonian 
and the doorway states, that is, the coupling coefficients occuring in 
the expansion of true eigenstates in the simple model basis.
Recently, the distribution of the maximum coupling coefficients was
introduced as a new, highly sensitive statistical observable. 
In the particularly important regime of weak interactions, this 
distribution is very well approximated by the fidelity distribution, defined as the 
distribution of the overlap between the doorway states with interaction and without
interaction. Using a random matrix model, we calculate the latter distribution 
exactly for regular and chaotic background states in the cases of 
preserved and fully broken time--reversal invariance. We also perform
numerical simulations and find excellent agreement with our analytical 
results.
\end{abstract}

\maketitle

\section{Introduction}
\label{sec1}

In open quantum systems,  strength function phenomena~\cite{boh69} 
give structural information about the system itself and about the excitation
mechanism. Here, we address statistical features of the doorway mechanism
which can be defined as follows: there are one or several somehow
``distinct'' and ``simple'' excitations whose amplitudes are spread over
many ``complicated'' states. In a many--body system, collective excitations
are often distinct, because all or large groups of particles move in a coherent
fashion. As compared to the complexity of the other, non--collective
excitations, these states can be interpreted in the framework of a ted in the framework of a 
simple, typically semiclassical, picture. The distinct states act as ``doorways''
to the background of the complicated states~\cite{boh69, sok97}. Mostly, 
the statistical features of the latter are chaotic. The strength function has
Breit--Wigner shape, largely independent of the statistics of the background
states. The width characterizing the Breit--Wigner strength function is 
referred to as spreading width~\cite{boh69}.

The doorway mechanism is found in a rich variety of systems, comprising atoms
and molecules \cite{her73}, as well as atomic clusters, quantum dots and, more generally,
mesoscopic systems~\cite{guh90,kaw00,hus00}. Nuclear
physics provides particularly beautiful and well--studied examples, such as
Isobaric Analog States and multipole Giant
Resonances~\cite{boh69,har86,zel96,abe04,shev04}.

What is a suitable theoretical interpretation of the Breit--Wigner shape? ---
Although the simple picture for the distinct excitations captures
the main physics, it is important to realize that these states
are not eigenstates of the real quantum Hamiltonian. Similarly, the statistical
models for the background states do not describe eigenstates either.
Thus, if we use the simple picture for the distinct states and the statistical
model for the background states as a basis of the Hilbert space, there must
be a non--vanishing interacting between these two classes of states.
Rediagonaliziation then yields proper eigenstates of the model Hamiltonian.
Averaging over the background states, one obtains the local density of
states around the energy of the distinct state or states in the simple picture.
The local density of states is once more of Lorentzian or Breit--Wigner shape
with a spreading width that is --- depending on the particular situation ---
closely related to or identical with the above mentioned spreading width
in the strength function. It can be viewed as a measure for the quality of the
simple picture describing the distinct states: the smaller the spreading width,
the closer is this picture to the physics reality. 
 
The strength of the interacting between the two classes of states uniquely
determines the spreading width and, equivalently, the size of the overlap
between the distinct state in the simple picture and the true eigenstates
of the model Hamiltonian. These overlaps are of course the coupling coefficients
when expanding the true eigenstates in the above mentioned basis. Recently, 
a new statistical observable was introduced: the distribution of the maximum
coupling coefficients~\cite{abe08}. The first two moments of this distribution
were already studied in Ref.~\cite{ver04}, but with assumptions not 
valid in our context. In Ref~\cite{abe08}, however, the full distribution 
is addressed. Importantly, its shape sensitively depends on the interaction 
strength. Moreover, it is an especially well--tailored
measure to investigate weak interactions.

Here, we present exact results for the distribution of the coupling coefficients
to a distinct state in the framework of a random matrix model.
In the particularly interesting regime of weak interactions,
this distribution coincides with the distribution of the maximum
coupling coefficients.

The article is organized as follows. After properly posing the problem
in Sec.~\ref{sec2}, we calculate the distribution exactly for regular
and chaotic background, respectively in Secs.~\ref{sec3} and~\ref{sec4}.
 We discuss our results in  Sec.~\ref{sec6}.

\section{Posing the Problem}
\label{sec2}

In Sec.~\ref{sec21}, we present the random matrix model for the doorway
mechanism. We introduce and define the distribution of the maximum 
coupling coefficient in Sec.~\ref{sec22}.

\subsection{Doorway Mechanism in a Random Matrix Model}
\label{sec21}

The model to be discussed here stems from nuclear physics~\cite{boh69} and
is also often used in other fields~\cite{guh98}. For the convenience of
the reader and to define our notation, we compile its salient features. As we
are aiming at a random matrix model, it is convenient to choose from the
beginning a proper basis of the full Hilbert space such that we can represent 
the Hilbert space operators by matrices. Introducing a cutoff, their dimension
is finite. Eventually this cutoff effect is removed by taking the matrix dimension
to infinity.  We nevertheless use the Dirac notation for the wave functions, 
even though they are finite--dimensional vectors.

The total Hamiltonian $H$ consists of three parts, the Hamiltonian
$H_s$ for the $K$ distinct states which become the doorway states,
the Hamiltonian $H_b$ describing the $N$ background states, where
$N$ will eventually be taken to infinity, and the interaction $V$ coupling 
the two classes of states. Hence, we have
\begin{eqnarray}
H&=& H_s+ H_b +V\nonumber\\
&=&\sum_{j=1}^K E_{sj}|s_j\rangle\langle s_j| + \sum_{\nu=1}^N E_{\nu}|b_\nu\rangle\langle b_\nu| + \sum_{j=1}^K\sum_{\nu=1}^N \Big( V_{j\nu} |s_j\rangle\langle b_\nu| + {\rm h.c.}\Big) \ .
\label{e1}
\end{eqnarray}
For the matrix elements of the interaction, we make the assumptions 
$\langle s_j|V|s_k\rangle=\langle b_\nu|V|b_\mu\rangle=0$ and
$\langle b_\nu|V|s_j\rangle=V_{\nu j}$ for any $j$, $k$, $\mu$, $\nu$. 
Often, there is only one relevant doorway state or the spacing between the
doorway states is much larger than their spreading widths. 
We focus on these cases and consider only one doorway state by setting 
$K=1$, $|s_1\rangle = |s\rangle$ and $V_{j\nu}=V_\nu$. 

The eigenequations for the uncoupled Hamiltonians are
\begin{equation}
H_s|s\rangle = E_s|s\rangle 
\qquad {\rm and} \qquad
H_b|b_\nu\rangle = E_\nu|b_\nu\rangle
\label{e2}
\end{equation}
Due to the interaction $V$ the doorway state  is not an
eigenstate of the Hamiltonian $H$. We denote the eigenstates of the 
full Hamiltonian $H$ by $|n\rangle$. The eigenequation to be solved is
\begin{equation}
H|n\rangle=E_n|n\rangle
\label{e3}
\end{equation}
Resembling the situation in most systems, we put the  doorway state 
$|s\rangle$ in the center of the background spectrum. It interacts with the
surrounding $N$ states. Without loss of generality, we may set $E_s=0$.
 
The exact eigenstate of $H$ which evolves from the doorway state 
in the presence of the interaction is referred to as $|0\rangle$. We expand
the $n$--th eigenstate of $H$ in the basis spanned by $|b_\nu\rangle$ 
and $|s\rangle$ as
\begin{equation}
|n\rangle \ =\ c_{ns} |s\rangle + \sum_{\nu=1}^N c_{n\nu}|b_\nu\rangle
\label{e4}
\end{equation}
where the coupling coefficient $c_{ns}$ is the overlap between the doorway
 state $|s\rangle$ in the non--exact picture for this distinct state and the 
$n$--th exact eigenstate $|n\rangle$ of the full Hamiltonian. We are interested in
the statistical features of these coupling coefficients.

We have to solve our model for $c_{ns}$. The action of the full
Hamiltonian $H$ on the eigenstate $|n\rangle$ yields on the one hand
\begin{equation}
H |n\rangle \ =\ \left(E_s c_{ns} + \sum_{\nu=1}^N V^*_\nu c_{n\nu}\right)|s\rangle + 
                    \sum_{\nu=1}^N\Big(c_{ns}V_{\nu}+ E_\nu c_{n\nu}\Big)|b_\nu\rangle \ .
\label{e5}
\end{equation}
On the other hand we have
\begin{equation}
H |n\rangle \ =\  E_n c_{ns} |s\rangle + E_n \sum_{\nu=1}^N c_{n\nu}|b_\nu\rangle \ .
\label{e6}
\end{equation}
Equating these two expressions, we find
\begin{equation}
c_{n\nu} \ =\ \frac{V_\nu}{E_n-E_\nu}c_{ns} \ ,
\label{e7}
\end{equation}
such that
\begin{equation}
|n\rangle \ =\ c_{ns} \left( |s\rangle + \sum_{\nu=1}^N \frac{V_\nu}{E_n-E_\nu} |b_\nu\rangle\right) \ .
\label{e8}
\end{equation}
Using the normalization of $|n\rangle$, we eventually arrive at
\begin{equation}
|c_{ns}|^2 \ =\ \left(1+\sum_{\nu=1}^N \frac{|V_\nu|^2}{(E_n-E_\nu)^2}\right)^{-1} \ ,
\label{e9}
\end{equation}
which is the desired expression for $c_{ns}$ in terms of the matrix 
elements of $H$. 

Formula~(\ref{e9}) holds for $0\leq n\leq N$. This expression is still
exact. However, since the eigenvalues of the full Hamiltonian $E_n$ depend 
on the coupling coefficents $V_\nu$, it is a complicated implicit expression. 
We proceed further by expanding the exact eigenvalues perturbatively in $V$,
\begin{eqnarray}
\label{perex}
E_{n} & = & E_{\nu(n)} + \frac{|V_{\nu(n)}|^2}{E_{\nu(n)}} 
+\sum_{\mu=1}^N\frac{|V_{\nu(n)}|^2|V_\mu|^2}{E_{\nu(n)}^3} +\ldots\qquad \ 
(1\leq n \leq N) \nonumber\\
E_0 &= &  E_s - \sum_{\nu=1}^N\frac{|V_\nu|^2}
{E_\nu} -\sum_{\mu=1}^N\sum_{\nu=1}^N\frac{|V_\nu|^2|V_\mu|^2}{E_\nu^3} \ ,
\end{eqnarray}
where the eigenstate $|n \rangle$ of the full Hamiltonian to eigenvalue 
$E_{n}$ has evolved from the eigenstate $|b_\nu(n)\rangle$ of the unperturbed 
Hamiltonian by adiabatically switching on the perturbation.
 
We now obtain a crucial simplification by setting $E_{n}= E_{\nu(n)}$, 
i.~e.~we keep only the leading order term in the perturbative expansion of 
Eq.~(\ref{perex}). To this approximation of $|c_{ns}|^2$ for $n\neq 0$ the sum in Eq.~(\ref{e9}) is completely 
dominated by the term $\nu = \nu(n)$, which actually diverges such that $|c_{ns}|^2 \approx 0$ to 
first order for all $n \neq 0$. The only overlap integral which remains 
finite is the overlap of the doorway state with itself $|c_{0s}|^2$. 
Here we set $E_0 \approx E_s$ $=0$ and no divergence occurs.
Using the approximation $E_{n}= E_{\nu(n)}$ we have therefore essentially 
singled out $|c_{0s}|^2$ as the only non--vanishing -- 
and thus inevitably maximum -- overlap integral of the perturbed 
eigenstates with the doorway state.

\subsection{Distribution of the Maximum Coupling Coefficient}
\label{sec22}

The new statistical observable introduced in Ref.~\cite{abe08}
is the distribution of the maximum 
\begin{equation}
c_{\rm max}= {\rm max}( |c_{ns}|, 0 \leq 
n\leq N)
\label{f1} 
\end{equation}
of the overlaps between the eigenstates of the full Hamiltonian and 
the distinct state, that is the doorway state $|s\rangle$. In order to obtain it, 
we have to average in a suitable way over the interaction matrix elements 
and over the Hamiltonian modeling the background states. For the
time being it suffices to denote this average by square brackets.
Later on we give a precise definition. Hence, the distribution in
question is given by
\begin{equation}
p_{\rm max}(c) \ =\ \langle \delta(c-c_{\rm max})\rangle \ .
\label{f2}
\end{equation} 
On the other hand, the distribution of overlap between the evolved doorway 
state and the unperturbed doorway state reads 
\begin{equation}
p_0(c) \ = \ \langle \delta(c-|c_{0s}|)\rangle \ .
\label{f3}
\end{equation}
Setting $E_{n}= E_{\nu(n)}$ amounts essentially to the approximation 
\begin{equation}
p_{\rm max}(c) \ \approx \ p_0(c) \ .
\label{f4}
\end{equation}
This approximation is certainly good for small interactions or, more precisely,
as long as the mean coupling strength is an order of magnitude smaller than 
the mean level density of the background states. Our 
numerical simulations will strongly corroborate this statement. Hence,
we focus on $p_0(c)$, which can be treated analytically. 

The statistics of the interaction matrix elements can only have minor 
impact on the resulting distribution. Hence, if not stated otherwise assume that the 
interaction matrix elements
are Gaussian distributed random variables. We have to distinguish two cases.
The total Hamiltonian $H$ can be time--reversal non--invariant
or time--reversal invariant, where we disregard spin degrees of freedom.
In the first case, labeled by the Dyson index $\beta=2$, the interaction 
matrix elements $V_\nu$ are complex variables, in the second case,
labeled $\beta=1$, they are real. Introducing the $N$--component vector
$V$, the corresponding distribution is
\begin{equation}
P_i(V) \ = \  \left(\frac{\beta}{2\pi v^2}\right)^{\beta N/2}
                \exp\left(-\frac{\beta}{2v^2}V^\dagger V\right) \ . 
\label{e9a}
\end{equation}
In addition to the behavior under time--reversal invariance, the statistical
properties of the Hamiltonian $H_b$, however, must strongly affect 
the distribution $p_0(c)$. Hence, we do not specify it yet. As is well known 
from Random Matrix Theory, the parameter governing the physics is
\begin{equation}
\lambda \ = \ \frac{\sqrt{\langle V^\dagger V}\rangle}{\sqrt{N}D} \ =\ \frac{v}{D} \ ,
\label{e9b}
\end{equation}
where $D$ is the mean level spacing of the background
states in the center of the band~\cite{boh69,guh98}. The distribution $P_i(V)$ 
is chosen such that $\lambda$ is independent of $\beta$. 

Technically, it is more convenient to work out the probability 
density $Q(u)$ of the random variable 
\begin{equation}
u \ = \ \frac{1}{c^2 } 
   \ = \ 1+\sum_{\nu=1}^N \frac{|V_\nu|^2}{E_\nu^2} \ .
\label{e10}
\end{equation}
The relation between the two distributions reads
\begin{equation}
p_0(c)  \ = \ Q(u) \left|\frac{du}{dc}\right|_{u=1/c^2}
                    \ =  \left.\frac{2}{c^3} Q(u)\right|_{u=1/c^2} \ . 
\label{e11}
\end{equation}
Thus, once $Q(u)$ is known, $p_0(c)$ follows immediately.

We now use our statistical assumption that the interaction matrix elements
$V_\nu$ are Gaussian distributed. We write the distribution $Q(u)$ in the form
\begin{equation}
Q(u) \ = \ \int d[V] P_i(V)
\left\langle\delta\left(u-1-\sum_{\nu=1}^N 
\frac{|V_\nu|^2}{E_\nu^{2}}\right)\right\rangle_N \ ,
\label{e12}
\end{equation}
where $d[V]$ is the product of the differentials of all independent variables
in $V$. The square brackets with index $N$ denote an average over the 
$N$ background
states, that is, over the Hamiltonian $H_b$. 
For the calculation of the averages, it is helpful to write the distribution $Q(u)$ 
as the Fourier transform
\begin{equation}
\label{genex1}
Q(u) \ = \ \frac{1}{2\pi}\int\limits_{-\infty}^{+\infty} 
                    dk \exp\left(ik(u-1)\right) R(k) 
\end{equation}
with the characteristic function 
\begin{equation}
R(k) \ = \ \int d[V] P_i(V)
\left\langle\exp\left(-ik\sum_{\nu=1}^N 
\frac{|V_\nu|^2}{E_\nu^{2}}\right)\right\rangle_N \ .
\label{genex1a}
\end{equation}
After rescaling $V_\nu = y_\nu/|E_\nu|$ we obtain the alternative expression
\begin{eqnarray}
R(k) &=& \sqrt{\frac{\beta}{2\pi v^2}}^{\beta N}\int d[y] \, 
             \exp\left(-ik y^\dagger y\right) \nonumber\\
     & & \qquad\qquad\qquad \left\langle |\det H_b|^\beta 
                  \exp\left(-\frac{\beta}{2v^2}y^\dagger H_b^2 y\right)
              \right\rangle_N \ ,
\label{genex}
\end{eqnarray} 
where the vector $y$ has real entries for $\beta=1$ and complex ones for
$\beta=2$, respectively. The infinitesimal volume element $d[y]$ is a product 
of the differentials of all independent entries of the vector $y$.  As a generating function $R$ is  normalized to $R(0)=1$.

In the sequel, we calculate the expressions (\ref{genex}) for generic choices 
of the Hamiltonian $H_b$ governing the dynamics of the background 
states.

\section{Regular Background}
\label{sec3}

The doorway state is embedded into a regular background, if the eigenvalues
$E_\nu$ of $H_b$ do not repel each other. The distribution of the
background Hamiltonian then factorizes according to
\begin{equation}
P_b(H_b) \ = \ \prod_{\nu=1}^N p_b(E_\nu) \ .
\label{e13}
\end{equation}
In order to keep the discussion most general we use for the interaction matrix elements a general factorizing distribution
\begin{equation}
P_i(V) \ = \ \prod_{\nu=1}^N p_i(V_\nu) \ ,
\label{e13a}
\end{equation} 
instead of the Gaussian distribution introduced before Eq.~(\ref{e9a}). We keep the reasonable, physically motivated assumption of statistical independence of the interaction matrix elements but relax the global orthogonal ($\beta=1$) or unitary ($\beta=2$) invariance of the interaction matrix elements, implicit in the measure Eq.~(\ref{e9a}). For complex coupling matrix elements we assume in addition that the distribution $p_i$ is $U(1)$ invariant $p_i(V)=p_i(|V|)$.  We asign to complex coupling matrix elements with this invariance the Dyson index $\beta=2$ and to real coupling matrix elements the Dyson index $\beta=1$. 
  
A straightforward calculation reveals that the characteristic 
function~(\ref{genex1a}) factorizes as well and becomes an 
$N$--th power of a single integral,
\begin{eqnarray}
R(k) &=& e^{N \ln r(k)}\\
r(k) &=& \int d^\beta[z]p_i(z)
            \int\limits_{-\infty}^{+\infty}dE
           p_b(E)\exp\left(-\frac{ik|z|^2}{E^2}\right)  .
\label{e14}
\end{eqnarray}
As we are interested in the local scale set by the mean level spacing
$D$ of the background states, the distribution $p_0(c)$ should 
not be sensitive to the particular choice of the distribution 
$p_b$, as long as it does not contain scales competing with
the mean level spacing $D$. The simplest choice is 
\begin{equation}
p_b(E) \ = \ \frac{1}{\sqrt{N}} 
\left\{
\begin{array}{ll}
 1 \ , \quad  & |E| \le \sqrt{N}/2 \cr
 0 \ , \quad  & |E| > \sqrt{N}/2 \ , 
\end{array} 
\right. 
\label{e15}
\end{equation}
where $D=1/\sqrt{N}$ and $\sqrt{N}=ND$ is the length of the background spectrum. The following calculation is similar to the one described in the Appendix B of Ref.~\cite{mel95}. We perform the integral over the background distribution in Eq.~(\ref{e14}) 
\begin{eqnarray}
\int\limits_{-\infty}^{+\infty}dE p_b(E)\exp\left(\frac{-ik|z|^2}{E^2}\right)&=& 
\frac{2}{\sqrt{N}} \int\limits_{2/\sqrt{N}}^{\infty} du \frac{\exp\left(-ik|z|^2u^2\right)}{u^2}\nonumber\\
 &\stackrel{N\to\infty}{=}&\frac{2}{\sqrt{N}} \int\limits_{0}^{\infty} du \frac{\exp\left(-ik|z|^2u^2\right)}{u^2}\nonumber\\
 &=& 1+2|z|\sqrt{\frac{i\pi k}{N}}\ . 
\end{eqnarray}
In the last equation we used an integral identity of the Fresnel type  
 \begin{eqnarray}
\frac{id}{d\alpha} \int\limits_{0}^{\infty} du \frac{\exp\left(-i\alpha u^2\right)}{u^2}&=&
\sqrt{\frac{i\pi}{4 \alpha}} \ .
\end{eqnarray}
We find for  the characteristic function
\begin{eqnarray}\label{m1def}
R(k)  &=&  \exp\left( 2 \sqrt{i\pi k N} m_1 \right)\\
 m_1  &=&\int d^\beta[z]p_i(z) |z|\ .
\end{eqnarray}
We observe that the distribution of the interaction matrix elements $p_i$ enters only via the expectation value $m_1$ as defined in Eq.~(\ref{m1def}) and {\em not} via the second moment $m_2=v^2$. As pointed out after Eq.~(\ref{e9b}) $p_i$ is chosen such that the mean coupling strength, as defined through $v$, is independent of $\beta$. This means that $m_1$ in general is different for real and for complex coupling. We write $m_1= a_\beta v$, where now $a_\beta$ depends on Dyson's index and on the distribution $p_i$. For instance for the Gaussian distribution  we find 
\begin{equation}
a^{\rm (G)}_\beta \ = \  
\left\{
\begin{array}{llll}
 \sqrt{\frac{2}{\pi}} &\approx& 0.80     \ , \quad  & \beta=1 \cr
 \sqrt{\frac{\pi}{4}} &\approx& 0.89\ , \quad  & \beta=2 \ .
\end{array} 
\right. 
\label{e17a}
\end{equation}
Using the definition of $\lambda$ in Eq.~(\ref{e9b}) we finally find
\begin{equation}
R(k)  \ = \ \exp\left(-2a_\beta\lambda\sqrt{i\pi k}\right) \ .
\label{e17}
\end{equation} 
 The reader can easily convince
herself/himself that other reasonable choices for $p_b(E)$ such 
as a Gaussian distribution yield the same functional form as in Eq.~(\ref{e17}). The Fourier
transform~(\ref{genex1}) results in
\begin{equation}
Q(u) \ = \  \frac{a_\beta \lambda}{(u-1)^{3/2}}
                \exp\left(-\left(a_\beta\lambda\right)^2\frac{\pi}{u-1}\right) \ . 
\label{e18}
\end{equation}
Using the relation~(\ref{e11}), we eventually arrive at
\begin{equation}
p_0(c)  \ = \  \frac{2 a_\beta\lambda }{(1-c^2)^{3/2}}
 \exp\left(-\left(a_\beta\lambda\right)^2
  \frac{\pi c^2}{1-c^2}\right) \ .
\label{e19}
\end{equation}
As anticipated, the interaction strenght enters the distribution only via
the dimensionless ratio $\lambda=v/D$. The distribution $p_i$ enters via the the factor $a_\beta$ defined in Eq.~(\ref{m1def}). It is interesting to see that $a_\beta$ does not only depend on the distribution but also on the symmetry factor $\beta$. This means that for a regular background and for constant interaction strength $\lambda$ the distribution $p_0$ distinguishes between real interaction  and  complex interaction. As we will see in the following this does not happen for a chaotic background. Therefore here opens -- at least theoretically -- the possiblity to distinguish between a regular and a chaotic background dynamics through the doorway state. 

Assume we can experimentally manipulate the interaction between the Doorway state and the background such the interaction matrix elements change from real to complex, for instance by switching on a magnetic field. For fixed mean interaction strength and for a chaotic background dynamics the distribution $p_0$ will be invariant, whereas for a regular background it will change.    
\begin{figure}
\begin{center}
\epsfig{figure=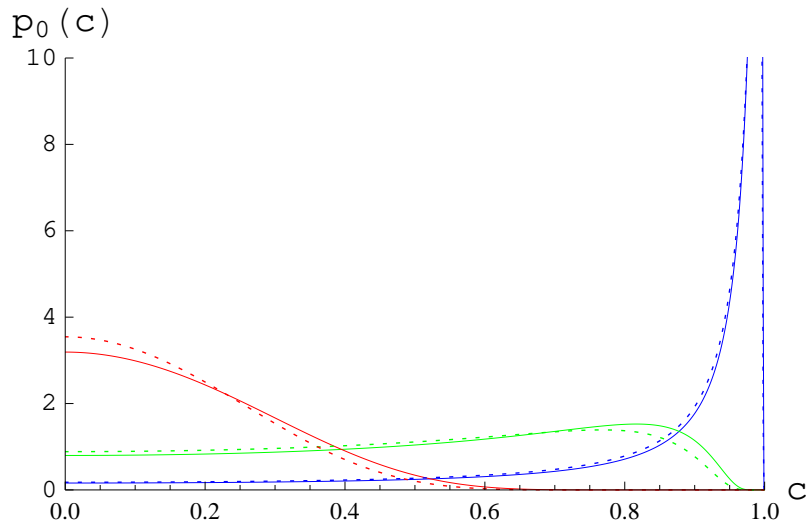}
\caption{\label{Figreg} Plot of the the distribution function $p_0(c)$ for a regular background for real interaction matrix elements (dotted) and for complex matrix element (full line) for three different values of the mean coupling strength 
$\lambda = 0.1$ (blue), $0.5$ (green), $2$ (red). The difference between real and complex coupling increases for strong coupling $\lambda$.}
\end{center}
\end{figure}

In Fig.~\ref{Figreg} $p_0$ is plotted for three different values of the mean coupling strength. We see that due to the numerical similarity of $a_1$ and $a_2$ the plots for real and for complex couplings are almost the same. The fact that for a Gaussian distribution  $a_1\approx a_2$ seems to be rather accidental. However for $p_i$ being a semicircle (SC) distribution  we find
$a^{\rm (SC)}_1$ $ \approx 0.85$ and  $a^{\rm (SC)}_2$ $\approx 0.93$ and for a uniform (U) distribution we find  
$a^{\rm (U)}_1$ $=$ $\sqrt{3}/2$ $\approx$ $0.87$ and $a^{\rm (U)}_2$ $=$ $\sqrt{8}/3$ $\approx 0.94$. In other words, for physically reasonable distribution functions $p_i$ we have $a_1 \approx a_2$, which encumbers the experimental possibility sketched above.

\section{Chaotic Background}
\label{sec4}

The dynamics of the background states is usually chaotic. The $N\times N$
Hamilton matrix $H_b$ modeling the background states has then 
to be chosen from a Gaussian random matrix 
ensemble. Again, we have to distinguish time--reversal invariant and
non--invariant systems, that is, the cases labeled by the Dyson parameters
$\beta=1$ and $\beta=2$, respectively. The Hamiltonian
$H_b$ is from the Gaussian orthogonal ensemble (GOE) 
for $\beta=1$ and from the Gaussian unitary ensemble (GUE) for $\beta=2$ with variance $w^2$ of the diagonal elements 
\begin{equation}
P_b(H_b) \ \sim \ \exp\left(-\frac{1}{2w^2}\tr H_b^2\right)
\label{e20}
\end{equation}
As already said, 
$y$ is an $N$--component vector which has real or complex entries
for $\beta=1$ and $\beta=2$, respectively.

In Sec.~\ref{sec41} we reformulate the problem in terms of matrix 
invariants. This enables us to introduce a handy supermatrix model
in Sec.~\ref{sec42}, which we then solve in Sec.~\ref{sec43}.

\subsection{Reformulation in a Rotation--Invariant Form}
\label{sec41}

At first sight, the best way of tackling the problem seems to start from
Eq.~(\ref{e12}) and to introduce $V_\nu=y_\nu/|E_\nu|$. The $y$ integration
is then simply an integration over a $N-1$ sphere in a real or
complex space. This leads to
\begin{eqnarray}
\label{eq0}
Q(u) &= & \frac{ \Omega_N}{2}  
                 \frac{(u-1)^{(\beta N-2)/2}}{(2v^2\pi/\beta)^{-\beta N/2}} 
                 \left< |\det H_b|^\beta e^{- \tr G H_b^2}\right>_N \ ,
\end{eqnarray}
where the $N\times N$ matrix 
\begin{equation}
G \ = \ \diag((u-1)\beta/2v^2,0,\ldots,0)
\label{e21}
\end{equation}
has rank one. Furthermore,
\begin{equation}
\Omega_N = \frac{2\sqrt{\pi}^{\beta N}}{\Gamma(\beta N/2)}
\label{e22}
\end{equation}
is the volume of the above mentioned sphere. Unfortunately, the remaining
ensemble average is only feasible for the GUE, we carry it out 
in~\ref{App1}. For the GOE, the calculation is hampered  by the modulus of 
the determinant. 

To adress the GOE case and to have a method that is capable of handling both
cases, GOE and GUE, in a unifying way, it turns out necessary to 
cast the ensemble average into an invariant form. To this end, we start from
Eq.~(\ref{genex}), perform the integration over the vector $y$ and obtain
\begin{eqnarray}
 \label{eq01}
R(k) &=& \left< \det(H_b^2+2iv^2k/\beta)^{-\beta/2}|\det H_b|^\beta \right>_N \ .
\end{eqnarray} 
Now we make the observation, that the $N\times N$--matrix average can be expressed as an $(N+1)\times (N+1)$--matrix average as follows
 \begin{eqnarray}
 \label{eq1}
&& \langle \det(H_b^2+2iv^2k/\beta)^{-\beta/2}|\det H_b|^\beta \rangle_N \ = \lim_{\epsilon\to 0} \ L_{N\beta} w^{\beta N}\sqrt{2\pi w^2}  
                \nonumber\\ 
                &&\qquad(\epsilon+2iv^2k/\beta)^{\beta/2} 
                 \left< \tr\delta(\widetilde{H})\det(\widetilde{H}^2+2iv^2k/\beta)^{-\beta/2} \right>_{N+1}\ , 
\end{eqnarray} 
where $\widetilde{H}$ is a $(N+1)\times (N+1)$ matrix and the ensemble average is over an $(N+1)\times (N+1)$--matrix GOE (GUE) ensemble. The derivation of Eq.~(\ref{eq1}), which is crucial for the calculation, is sketched in App.~\ref{App2}. On the right hand side, the inconvenient modulus of the determinant 
has disappeared. The combinatorial factor $L_{N\beta}$ is given by
 \begin{equation}
 \label{Lcon}
L_{N\beta} \ =\ \left\{
\begin{array}{cc}
\displaystyle \frac{2 \Gamma(1+(N+1)/2)}{\sqrt{\pi}(N+1)} \ , 
                 \quad& \beta = 1 \cr
N! \ , \quad & \beta =2 
\end{array} \right. \ .
 \end{equation} 
We express the determinant on the right hand side of Eq.~(\ref{eq1}) as 
a Gaussian integral over a real ($\beta =1$) or complex ($\beta=2$) $N+1$ vector $y$.
\begin{eqnarray}
\label{backw}
\det(H_b^2+2iv^2k/\beta+\epsilon)^{-\beta/2}& = & 
                                   \int \frac{d[y]}{\pi^{\beta/2}}\exp\left(-y^\dagger(H_b^2+2iv^2k/\beta+\epsilon)y\right)
\end{eqnarray}
and plug subsequently Eqs.~(\ref{eq01}),~(\ref{eq1}) and ~(\ref{backw}) into Eq.~(\ref{genex1}) and write the integral over the $N+1$ vector $y$ in radial coordinates. We find  
\begin{eqnarray}
Q(u) &= & \frac{ L_{N\beta}\Omega_{N+1}}{2\pi^{\beta/2}}\left(\frac{\beta w^2}{2v^2 \pi}\right)^{\beta N/2} \nonumber\\ 
     && \lim_{\epsilon\to 0} \int\limits_0^\infty dx x^{\beta(N+1)/2-1}\widetilde{F}_{N+1}\left(1+\frac{\beta xw^2}{v^2}\right)  \nonumber\\
      && \int\limits_{-\infty}^\infty \frac{dk (\epsilon+ik)^{\beta/2}}{2\pi} e^{ik(u-1-x)} \ ,
\end{eqnarray}
where we introduced the function
\begin{equation}
\label{fdef}
\widetilde{F}_{N+1}(g) \ =\  \sqrt{2\pi w^2} \left< \tr\delta(\widetilde{H}) e^{-\frac{1}{2w^2} \tr \widetilde{G} \widetilde{H}^2 }\right>_{N+1}  
\end{equation}
with the $(N+1)\times (N+1)$ matrix $\widetilde{G} = \diag(g-1,0,\ldots,0)$. 

After combining the various constants we arrive at
\begin{eqnarray}
\label{qgen}
Q(u) &= & \frac{1}{\Gamma(\beta/2)}\left(\frac{\beta w^2}{2v^2}\right)^{\beta N/2}\frac{d^{\beta/2}}{d(u-1)^{\beta/2}}\nonumber\\ 
     && \int\limits_0^\infty dx x^{\beta(N+1)/2-1}  \widetilde{F}_{N+1}\left(1+\frac{\beta xw^2}{v^2}\right)
       \delta(u-1-x) \ ,
\end{eqnarray}
where we introduced formally the fractional derivative
\begin{equation}
\frac{d^{\beta/2}}{dx^{\beta/2}}\delta(x) \ =\ \lim_{\epsilon\to 0} \frac{1}{2\pi}\int\limits_{-\infty}^\infty dk (\epsilon+ik)^{\beta/2} e^{ikx} \ .
\end{equation}
We can evaluate the fractional distribution
\begin{eqnarray}
\frac{d^{\beta/2}}{dx^{\beta/2}}\delta(x) & = & \frac{1}{\pi}{\rm Re} 
                           \lim_{\epsilon\to 0} 
                           \frac{e^{+\frac{i\pi\beta}{4}} \Gamma\left(\frac{\beta}{2}+1\right)}
                            {\left(\epsilon-ix\right)^{\beta/2+1}} \nonumber\\
                   &=& \frac{2}{\pi\beta}\frac{d}{dx}{\rm Im} 
                           \lim_{\epsilon\to 0} 
                           \frac{e^{+\frac{i\pi\beta}{4}} \Gamma\left(\frac{\beta}{2}+1\right)}
                            {\left(\epsilon-ix\right)^{\beta/2}} 
\end{eqnarray}
for arbitrary $\beta$. For $\beta=1,2$ we obtain
\begin{eqnarray}
\frac{d^{\beta/2}}{dx^{\beta/2}}\delta(x)  &=& \left\{
\begin{array}{ll}{\displaystyle \frac{d}{d x} \frac{2 \Gamma(3/2)}{\pi \sqrt{x}} \Theta(x)}&  \beta =1\cr
{\displaystyle \frac{d}{d x} \delta(x)}&  \beta=2 . \cr
\end{array} 
\right. 
\end{eqnarray}
Here we see that the GOE case is more complicated than the GUE case.

For the GUE the fractional derivative disappears and we find without further problems
\begin{eqnarray}
\label{eqgue}
Q(u) &= &  \frac{d}{du} \left(\frac{\beta (u-1)w^2}{2 v^2}\right)^{N} 
                 \widetilde{F}_{N+1}\left(1+\frac{\beta (u-1)w^2}{v^2}\right)  \ .
\end{eqnarray}
For the GOE we obtain an integral expression
\begin{eqnarray}
\label{eqgoe}
Q(u) &= & \frac{1}{\pi} \frac{d}{du} \left(\frac{\beta w^2}{2 v^2}\right)^{N/2}\int\limits_0^{(u-1)}\frac{x^{\frac{N-1}{2}}}{\sqrt{u-1-x}}
\widetilde{F}_{N+1}\left(1+\frac{\beta xw^2}{v^2}\right) dx
\end{eqnarray}
The remaining task is in both cases (GOE and GUE) the calculation of $\widetilde{F}_{N+1}(g)$. 


\subsection{Mapping onto a Supermatrix Model}
\label{sec42}

Using 
\begin{eqnarray}
\tr \delta(H) &=& \frac{1}{\pi}\im \tr\frac{1}{H-i\epsilon}\nonumber\\
              &=&  \left.\frac{1}{2 \pi}\im\frac{{\rm d}}{{\rm d}j}\frac{\det(H+j)}{\det(H-j-i\epsilon)}\right|_{j=0} 
\end{eqnarray}
the ensemble average $\widetilde{F}_{N+1}(g)$ defined in Eq.~(\ref{fdef}) can be expressed via standard techniques 
as a supersymmetric matrix integral
\begin{eqnarray}
\label{fmat}
\widetilde{F}_{N+1}(g)&=&  \sqrt{\frac{w^2}{2 \pi}}\im\frac{{\rm d}}{{\rm d} j}\frac{1}{\sqrt{g}}
                          \left(\frac{2}{1+g}\right)^{\beta N/2}\nonumber\\
           &&  \int  d[\tau] 
               \exp\left(-\frac{(1-g)^2}{8gw^2}\Str\left(\tau+\frac{1+g}{1-g}J\right)^2\right) \Sdet^{-1}(\tau^-+J)\nonumber\\
           &&\left. \int d[\sigma] \exp\left(-\frac{1}{2w^2}\Str\sigma^2\right) \Sdet^{-\beta N/2}(\sigma^-+\tau)\right|_{j=0} \ .
\end{eqnarray}
Here $\sigma$ and $\tau$ are $2\times 2$ (GUE) respectively $4\times 4$ (GOE) supermatrices of the form
\begin{eqnarray}
\left(\begin{array}{cc}
a_1& \lambda_1^*\cr
\lambda_1& ia_2	
\end{array}\right)&,\quad & {\rm GUE}\nonumber\\ 
\left(\begin{array}{cccc}
a_1&a_2& \lambda_1^*&-\lambda_1\cr
a_2&a_3& \lambda_2^*&-\lambda_2\cr	
\lambda_1&\lambda_2&ia_4&0\cr
\lambda_1^*&\lambda_2^*&0&ia_4
\end{array}\right)&,\quad& {\rm GOE} \ .
\end{eqnarray}
The matrix entries in latin letters denote real commuting integration variables. The matrix entries in greek letters denote complex anticommuting integration variables. The infinitesimal volume elements $d[\tau]$ and $d[\sigma]$ are products of the differentials of all independent integration variables. The integration domain of the real commuting variables is the real axis. The matrix $J$ is a $2\times2$ (GUE) or a $4\times4$ (GOE) diagonal supermatrix with entries $J = \diag(j,-j)$ (GUE) and  $J = \diag(j,j,-j,-j)$ (GOE). Due to the broken rotation invariance of the original matrix model (\ref{e1}) the resulting supersymmetric representation (\ref{fmat}) is a two--matrix model.
 
We wish to evaluate the $\sigma$--integral by a saddle--point approximation and to calculate the $\tau$ integral exactly afterwards. 
It is well known \cite{ver85} that the $\sigma$ integral $K_N(\tau)$ yields for large $N$ in the saddle--point approximation
 \begin{eqnarray}
\label{saddle}
K_N(\tau)&=& \int d[\sigma] \exp\left(-\frac{1}{2w^2}\Str\sigma^2\right) \Sdet^{-\beta N/2}(\sigma^-+\tau) \nonumber\\
         &\simeq& \exp\left(-\frac{1}{2w^2}\Str \tau^2+\frac{i\beta\pi\Str\tau}{2D}\right) + {\cal O}\left(\frac{1}{N}\right)\ ,
\end{eqnarray}
where $D= \sqrt{\beta\pi^2w^2/(2 N)}$ is the mean level spacing in the center of the band. However this approximation is only valid, if $\Str \tau$ itself is of order of the mean level spacing. Since the integration domain of $\tau$ is the whole real axis, this is not automatically guaranteed. A necessary condition is that the variance of the Gaussian in the second line of Eq.~(\ref{fmat}) is itself of order of the mean level spacing, i.~e. the $\tau$ integral in Eq.~(\ref{fmat}) is essentially localised to a small window of width $D$ around zero. Consequently we must require
\begin{equation}
\frac{(1-g)^2}{8gw^2}\simeq  N+{\cal O}(1) \ . 
\end{equation}
And therefore $g$ should scale as $N$ for large $N$. 
 The dimensionless coupling strength is given by $\lambda = v/D$. Since $u$ it is of order one, we obtain for $g$ in the GUE case
\begin{eqnarray}
\label{scaling}
g & = & 1 + \frac{2(u-1)}{\pi^2\lambda^2} N\nonumber\\
  & \simeq &  \frac{2(u-1)}{\pi^2\lambda^2} N \ .
\end{eqnarray}
Fortunately this is exactly the scaling behaviour we need to apply the saddle--point approximation. In the GOE case Eq.~(\ref{scaling}) holds in any Intervall  $\omega_c<x<(u-1)$, where $\omega_c$ is an infrared cutoff in the integral, which is small compared to one but large compared to the mean level spacing.  i.e.~ in the limit $N\to \infty$ in the whole integration domain of the $x$--integral in  Eq.~(\ref{eqgoe}). In conclusion we can apply the approximation (\ref{saddle}) both in the GUE as well as in the GOE case.

\subsection{Remaining Matrix Integration and Final Result}
\label{sec43}

Plugging Eq.~(\ref{saddle}) into Eq.~(\ref{fmat}) we obtain after a simple shift
\begin{eqnarray}
\label{fmata}
&&\widetilde{F}_{N+1}(g)\ =\  \sqrt{\frac{w^2}{2 \pi}}\im\frac{{\rm d}}{{\rm d} j}\frac{1}{\sqrt{g}}
                          \left(\frac{2}{1+g}\right)^{\beta N/2} 
                          \exp\left(\frac{2i\pi j}{D}\right)\nonumber\\
           &&\qquad  \left.\int  d[\tau] 
               \exp\left(-\frac{g}{2w^2}\Str\left(\tau-\frac{i\beta\pi w^2}{Dg}-J \right)^2\right) 
            \Sdet^{-\beta/2}\tau^-  \right|_{j=0} \ ,
\end{eqnarray} 
where we also employed that $g\simeq N$ $\gg 1$. 
Now the derivative with respect to the source term can be performed
\begin{eqnarray}
\label{fmat1}
\widetilde{F}_{N+1}(g)&=&  \frac{\sqrt{2\pi w^2}}{D}\frac{1}{\sqrt{g}}\left(\frac{2}{1+g}\right)^{\beta N/2}\nonumber\\
                             &&   \left(1 +
            \frac{D \sqrt{g}}{\pi}\im \left< \tr \frac{1}{H+\frac{\beta i\pi w^2}{D\sqrt{g}}}\right>_{1}\right)\ ,
\end{eqnarray} 
where we used the identity
\begin{eqnarray}
 \left<\tr\frac{1}{H+z}\right>_N
              &=&  \left.\frac{1}{2}\frac{{\rm d}}{{\rm d}j} \int d[\tau] 
               e^{-\frac{1}{2w^2}\Str\left(\tau+ z -J \right)^2} \Sdet^{-\beta N/2}\tau^- \right|_{j=0}\ . 
\end{eqnarray}
for a complex number $z$ with negative imaginary part.
The remaining average can be calculated by employing techniques of standard analysis. We use
\begin{equation}
{\rm Im}\frac{1}{2\pi}\int\limits_{-\infty}^{\infty}dx 
\frac{\exp\left(-x^2/2\right)}{x-ir} \ = \ {\rm sgn}(r)\sqrt{\frac{\pi}{2}}
 \exp\left(\frac{r^2}{2}\right){\rm erfc}\left(\sqrt{\frac{|r|}{2}}\right) \ ,
\end{equation}
which holds for $r \in {\mathbb R}$. We obtain
\begin{eqnarray}
\label{fmat2}
\im \left< \tr \frac{1}{H+\frac{\beta i\pi w^2}{D\sqrt{g}}}\right>_{1}&=& 
                         \sqrt{\frac{\pi}{2w^2}}\exp\left(\frac{\beta N}{g}\right){\rm erfc}\left(\sqrt{\frac{\beta N}{g}}\right)\ .
\end{eqnarray} 
Finally we obtain
\begin{eqnarray}
\label{fmat3}
\widetilde{F}_{N+1}(g)&=&  \sqrt{\frac{2\pi w^2}{gD^2}}
                           \left(\frac{2}{1+g}\right)^{\beta N/2}\nonumber\\
                     &&\quad 
                     \left(1+ \frac{\beta}{2}\sqrt{\frac{\pi g}{\beta N}} 
           \exp\left(\frac{\beta N}{g}\right){\rm erfc}\left(\sqrt{\frac{\beta N}{g}}\right)\right) \ .
\end{eqnarray}
This result simplifies considerable when we take into account the scaling behavior (\ref{scaling}) of $g$. In the large $N$ limit we can write
\begin{eqnarray}
\label{fmat4}
&&\lim_{N\to\infty}\left(\frac{\beta x w^2}{2 v^2}\right)^{\beta N/2} \widetilde{F}_{N+1}\left(1+\frac{\beta xw^2}{v^2}\right)\ =\ \nonumber\\
 &&\qquad \sqrt{\frac{2\pi\lambda^2}{\beta x}}\exp\left(-\frac{\beta(\pi\lambda)^2}{2 x}\right)+
            {\rm erfc}\left({\sqrt{\frac{\beta(\pi\lambda)^2}{2 x}}}\right)
\end{eqnarray} 
This can now be plugged into Eq.~(\ref{qgen}) to obtain expressions for $Q(u)$ on the scale of the mean level spacing.
For the GUE we find straightforwardly
\begin{eqnarray}
\label{gueresult}
Q(u)&=& \sqrt{\frac{\pi\lambda^2}{4(u-1)^3}} \exp\left(-\frac{\pi^2\lambda^2}{(u-1)}\right)\left(1+\frac{2\pi^2\lambda^2}{u-1}\right)\nonumber\\
p_0(c) &=& \sqrt{\frac{\pi\lambda^2}{(1-c^2)^3}}\exp\left(-\frac{\pi^2\lambda^2c^2}{(1-c^2)}\right)
 \left(1+\frac{2\pi^2\lambda^2c^2}{1-c^2}\right)
\end{eqnarray}
For the GOE we are left with an integral expression for $Q(u)$
\begin{eqnarray}
Q(u) &=&  \sqrt{\frac{\pi^3\lambda^6}{2(u-1)^5}}\int\limits_0^1\frac{dx}{\sqrt{1-x}x^2}\exp\left(-\frac{\pi^2\lambda^2}{2(u-1)x}\right)
\end{eqnarray}
The integral can be evaluated further and be expressed in terms of standard special functions. Finally we arrive at
\begin{eqnarray}
\label{goeresult}
Q(u) &=& \sqrt{\frac{\pi^3\lambda^6}{8(u-1)^5}}\exp\left(-\frac{\pi^2\lambda^2}{4(u-1)}\right)\nonumber\\
      &&\qquad\left[K_0\left(\frac{\pi^2\lambda^2}{4(u-1)}\right)+
              K_1\left(\frac{\pi^2\lambda^2}{4(u-1)}\right)\right]\nonumber\\
p_0(c) &=& \sqrt{\frac{\pi^3\lambda^6 c^4}{2(1-c^2)^5}}
                  \exp\left(-\frac{\pi^2\lambda^2 c^2}{4(1-c^2)}\right)\nonumber\\
&& \qquad \left[K_0\left(\frac{\pi^2\lambda^2 c^2}{4(1-c^2)}\right)+
      K_1\left(\frac{\pi^2\lambda^2 c^2}{4(1-c^2)}\right)\right] \ ,
\end{eqnarray}
where $K_{n}$ is the modified Bessel function of the second kind of order $n$.
\begin{figure}
\begin{center}
\epsfig{figure=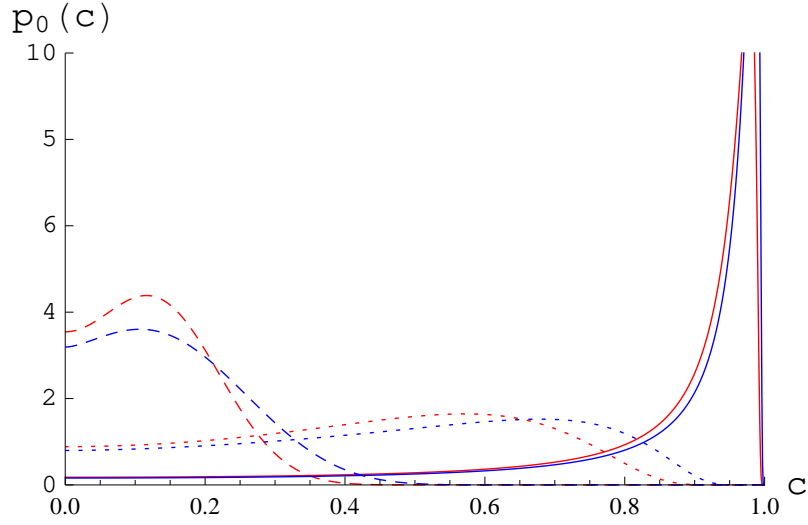}
\caption{\label{myres}Figure showing the plots of the analytical results for the GOE (blue curves) and for the GUE (red curves) for three different coupling strength ($\lambda = v/D$) $\lambda=0.1$ (thick curves), $\lambda=0.5$ (dashed curves) and $\lambda=2$ (dotted curves)}
\end{center}
\end{figure}
In the Fig.~(\ref{myres}) the distributions $p_0(c)$ of Eq.~(\ref{goeresult}) for the GOE (blue curves) and of Eq.~(\ref{gueresult}) for the GUE (red curves) are plotted for the values $\lambda = 0.1$, $0.5$, $2$ of the mean coupling strength $\lambda$. We see that for small $\lambda$ there is only a minor difference between GUE and GOE background.

\subsection{Comparison}
\label{sec5}

\begin{figure}
\begin{center}
\epsfig{figure=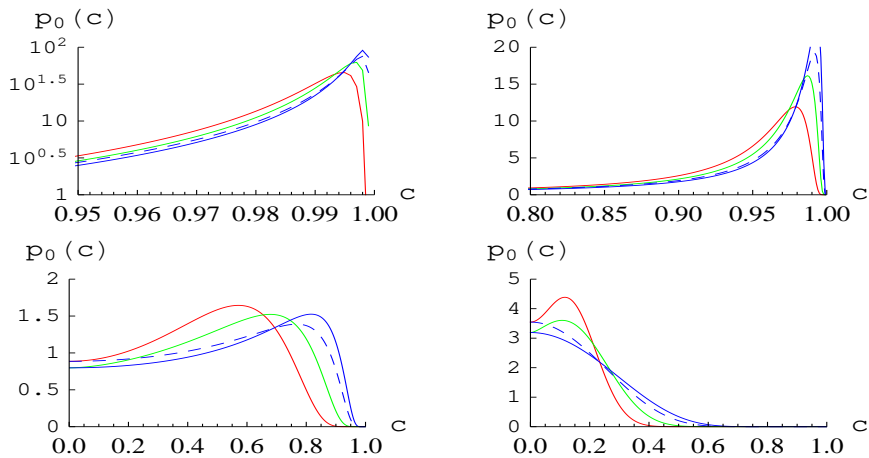, height=6cm, width=12cm}
\caption{\label{figcom} Four figures showing the analytical results 1) for a complex coupling of the doorway state to a GUE background (full red line) 2) for a real coupling to a GOE background (full green line), 3) for a complex coupling to a regular background (dashed blue line) and 4) for a real coupling to a regular background (full blue line) for four different coupling strength $\lambda = v/D$: $\lambda=0.05$ (upper left panel), $\lambda=0.1$ (upper right panel), $\lambda=0.5$ (lower left panel) and $\lambda= 2$ (lower right panel).}
\end{center}
\end{figure}

In Fig.~\ref{figcom} the distribution function of the overlap integral $|\langle 0|s\rangle|$ of the evolved doorway state with the unperturbed doorway state is plotted for four different coupling strengths $\lambda = 0.05$, $0.1$, $0.5$ and $2$ and for all types of couplings and background complexities considered. These are: 1) complex coupling to a regular background (dashed blue line), 2) real coupling to a regular background (full blue line), 3) complex coupling to a GUE background (full red line) and 4) real coupling to a GOE background (full green line). As a general trend mixing with the background is strongest for a complex coupling to a GUE background and weakest for real coupling to Poissonian background. However the difference of the distributions for different background complexities is rather small. This suggests a certain degree of universality of the curves. One might choose other ensembles for the background Hamiltonian, as for instance semi--Poisson\cite{bog06} or  transition ensembles. However we expect that for these ensembles which lie between the two extreme cases, GUE and Poissonian, their corresponding distributions will also lie in the channel between the full red line (GUE) and the full blue line (Poissonian with real coupling). For the most interesting case of small $\lambda$ this channel is small. 
   
On the other hand the distributions are highly sensitive with respect to a change in the coupling strength $\lambda$. 

\begin{figure}
\begin{center}
\epsfig{figure=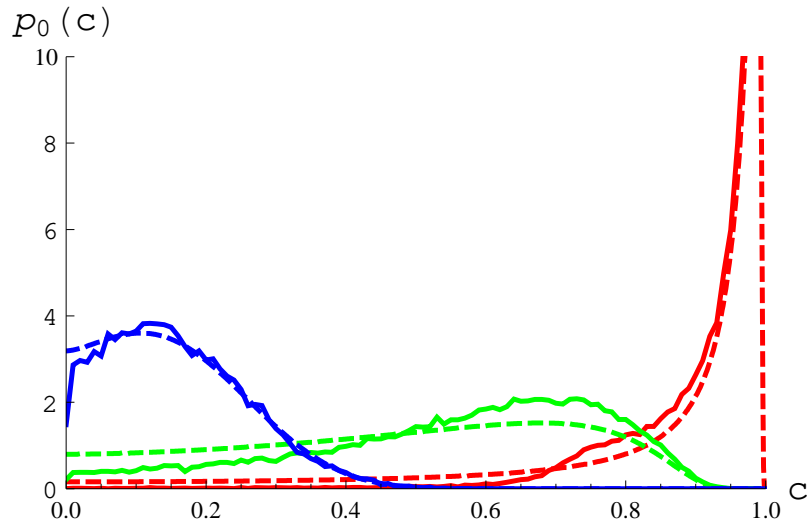}
\caption{\label{fignum} Comparison of the analytical curves (dashed lines) obtained from Eq.~(\ref{goeresult}) with Monte Carlo simulations (full lines) for $\lambda=0.1$ (red), $\lambda=0.5$ (green) and $\lambda=2$ (blue) for a GOE background and real coupling coefficients.}
\end{center}
\end{figure}

In Fig.~\ref{fignum} we compare the curves for $p_0(c=|\langle 0|s\rangle|)$ obtained from the analytical results (in this case  from Eq.~(\ref{goeresult})) with Monte Carlo simulations in the case of a real coupling to a GOE background. The figure shows $p_0(c=|\langle 0|s\rangle|)$  for three values of the coupling strength $\lambda = 0.1$, $0.5$ and $2$. We see fairly good agreement for all three values, even for the strong coupling value $\lambda=2$. This shows that the approximation implied in Eq.~(\ref{perex}) and thereafter, is justified far beyond the perturbative regime. 

\section{Discussion}
\label{sec6}

The distribution of the maximum coupling coefficients in the doorway
mechanism has been introduced as a new statistical observable. 
These coupling coefficients, that is, the overlaps
between the eigenstates of the full Hamiltonian and the doorway state
are not always at the disposal. However, in situations where they are 
accessible, this distribution provides a highly sensitive measure
for the interaction strength. Of particular interest is the regime of
weak interactions. In this regime, the distribution of the maximum 
coupling coefficients is very well approximated by the distribution 
of the overlap between the evolved doorway state and the unperturbed 
doorway state. While calculating the former seems unfeasible at present, 
we calculated the latter exactly for regular and chaotic background states
in the cases of preserved and fully broken time--reversal invariance.
We performed our calculations in the framework of Random Matrix Theory
which is well--known to provide reliable models for regular and chaotic
systems. We also carried out numerical simulations which fully confirm 
our analytical results. 

Our exact calculations are of general interest for matrix models. We 
managed to reformulate a problem with breaking of rotation invariance
in the space of $N\times N$ random matrices in terms of a  rotation
invariant problem involving  $(N+1)\times (N+1)$ random matrices.
This made it possible to map the matrix model in ordinary space onto
a matrix model in superspace, which we solved by a saddle point approximation 
in the limit of infinite level number. Remarkably, the supermatrix model 
is in the class of two--matrix models which show up in a large 
variety of situations.

\ack We thank P.~Mello, R. Sch{\"a}fer, T.~H.~Seligman and H.~J. Sommers for useful discussions. 
Two of us (HK and TG) acknowledge support from Deutsche Forschungsgemeinschaft (DFG) within
Sonderforschungsbereich Transregio 12 ``Symmetries and Universality in
Mesoscopic Systems". HK is grateful for financial support from DFG, with grant 
No. Ko 3538/1-1 and 3538/1-2. Part of this work was done during the workshop/conference 
at the Centro Internacional de Ciencias (C.I.C) in Cuernavaca, Mexiko in March 2nd to 6th 2009.

\appendix

\section{GUE Background with Finite Level Number}
\label{App1}

We  define the average
\begin{equation}
\langle\ldots\rangle_{gN} \ =\ A_N\int d[H] (\ldots) e^{-\frac{1}{2w^2}\tr G^\prime H^2} \ ,
\end{equation}
where the diagonal $N\times N$ matrix $G$ is defined as $G= \diag(g^\prime,1,\ldots,1)$. Her $g^\prime$ is related to the parameter of the main text as $g^\prime=1+ 2(u-1)w^2/v^2$. The integration is over the set of all Hermitean $N\times N$ matrices, i.~e. over the GUE ensemble. The normalisation $A_N$ is chosen such that
 $\langle 1\rangle_{1N} =1$. The task is to calculate the average 
 \begin{equation}
 F_N(g^\prime)=\langle\det H^2\rangle_{g^\prime N} \ .
 \end{equation}
A Laplace expansion of the determinant yields
\begin{equation}
 F_N(g^\prime)=\sum_{\omega,\omega^\prime\in S_N} (-1)^{\sgn(\omega)+\sgn(\omega^\prime)}\langle \prod_{n=1}^N H_{n\omega(n)}H_{\omega^\prime(n)n}\rangle_{g^\prime N}\ ,
\end{equation}
where $S_N$ is the permutation group. Obviously only terms $\omega = \omega^\prime$ contribute. 
\begin{equation}
 F_N(g^\prime)\ =\ \sum_{\omega\in S_N}\langle \prod_{n=1}^N |H_{n\omega(n)}|^2\rangle_{g^\prime N}\ ,
\end{equation}
It is useful to expand the remaining sum in cycles involving the index $1$. For indices $k_n >1$ and $k_n\neq k_m$ we define the cycles ${\cal C}^{(1)}_n$ as
\begin{eqnarray}
{\cal C}^{(1)}_n(k_1,\ldots k_n) & = & |H_{1k_1}|^2|H_{k_1k_2}|^2\ldots |H_{k_{n-1}k_n}|^2|H_{k_n1}|^2
\nonumber\\
{\cal C}^{(1)}_0  & = & |H_{11}|^2
\end{eqnarray}
Since the indices $1,k_1,\ldots k_n$ do not appear in the remainder of the product, we can integrate over the remainder separately. This yields
\begin{equation}
 F_N(g^\prime) =\ \sum_{n=0}^{N-1}\sum_{k_1\neq k_2\ldots k_n} \langle{\cal C}^{(1)}_n(k_1,\ldots,k_n)\rangle_{gN} F_{N-n-1}(1) 
\end{equation}
The average over the cycles is simple as well. Only terms involving the index $1$ yield a factor different from $w^2$.
We ontain
\begin{eqnarray}
\langle{\cal C}^{(1)}_0\rangle &=& \left(\frac{2}{g^\prime+1}\right)^{N-1}\frac{w^2}{g^{\prime 3/2}}\nonumber\\
\langle{\cal C}^{(1)}_1(k_1)\rangle &=& \left(\frac{2}{g^\prime+1}\right)^{N+1} \frac{2w^4}{\sqrt{g^\prime}}\nonumber\\
\langle{\cal C}^{(1)}_n(k_1,\ldots k_n)\rangle &=& \left(\frac{2}{g^\prime+1}\right)^{N+1} \frac{w^{2(n+1)}}{\sqrt{g^\prime}}\ ,\quad n> 1
\end{eqnarray}
The averages are independent of the indices $k_n$. The sum over the indices yields the combinatorial factor
$(N-1)!/(N-1-n)!$. Altogether we obtain
\begin{eqnarray}
\label{app2}
 F_N(g^\prime) & = & \frac{1}{\sqrt{g^\prime}} \left(\frac{2}{g^\prime+1}\right)^{N-1}\left(
                  \frac{w^2}{g^\prime}F_{N-1}(1)+ \right.\nonumber\\
&&\qquad 2(N-1)w^4\left(\frac{2}{g^\prime+1}\right)^{2} F_{N-2}(1)+\nonumber\\
&&\left.\left(\frac{2}{g^\prime+1}\right)^{2}\sum_{n=2}^{N-1}w^{2(n+1)}\frac{(N-1)!}{(N-1-n)!} F_{N-n-1}(1)\right)\ . 
\end{eqnarray}
Evaluating this equation for $g^\prime=1$ allows us to replace the sum. After some further simple manipulations we finally obtain
\begin{eqnarray}
 F_N(g^\prime) & = & \frac{1}{\sqrt{g^\prime}} \left(\frac{2}{g^\prime+1}\right)^{N+1}\left( F_N(1)+\frac{w^2(g^\prime-1)^2}{4g^\prime} F_{N-1}(1)\right)\ , 
\end{eqnarray}
 which is almost our final result. The remaining task is to evaluate the $g^\prime$ independent constant $F_N(1)$. 
 This is facilitated by the observation that 
 \begin{equation}
 \label{app1}
 F_N(1) \ =\ \sqrt{\pi}(\beta w^2)^N N! K_{N+1}(0,0) \,
 \end{equation}
 where $K_{N+1}(x,y)=\sum_{n=0}^N\phi_n(x)\phi(y)$ is the standard GUE kernel as defined in Eq.~(6.2.10) of Metha's book (third edition). The 
 \begin{equation}
 \phi_n(x)\ =\ (2^nn!\sqrt{\pi})^{-1/2}\exp(-x^2/2) H_n(x) 
 \end{equation}
 are oscillator wave functions and $H_n$ is the $n$--th Hermite polynomial. The constant $K_{N}(0,0)$ can also be evaluated
 \begin{equation}
 K_{N}(0,0) \ =\ \frac{1}{\sqrt{\pi}}\frac{N!}{2^{N-1}}\left\{
\begin{array}{ll}
	[(N-1)/2)!]^{-2} & N \ {\rm odd}\cr
	[(N-2)/2)!]^{-2} & N \ {\rm  even \ .}\cr
\end{array} \right.
 \end{equation}
Of course $K_{N}(0,0)$ is but the inverse level spacing at the center of the semicircle. Therefore  $\lim_{N\to\infty}K_{N}(0,0)/\sqrt{2N} = 1/\pi$. We obtain our final result
\begin{eqnarray}
 F_N(g^\prime) & = & (2w^2)^NN! K_{N+1}(0,0) \sqrt{\frac{\pi}{g^\prime}} 
    \left(\frac{2}{g^\prime +1}\right)^{N+1}\left(\frac{(g^\prime -1)^2}{4 g^\prime N}+ c_N\right)\ , 
\end{eqnarray} 
where 
 \begin{equation}
 c_{N} \ =\ \left\{
\begin{array}{ll}
	1 & N \ {\rm odd}\cr
	1+ 1/N& N \ {\rm  even}\cr
\end{array} \right.
 \end{equation} 
The even--odd difference disappears in the large $N$ limit. In the following we set $c_N=1$. Using Eq.~(\ref{eq0}) we find for $Q(u)$
\begin{eqnarray}
Q(u) &= &  \left(\frac{w^{2}}{ v^2}\right)^{N}K_{N+1}(0,0) (u-1)^{N-1} \nonumber\\
    && \quad\sqrt{\frac{\pi}{g^\prime}} \left(\frac{2}{g^\prime +1}\right)^{N+1}
   \left(\frac{(g^\prime -1)^2} {4 g^\prime} + N \right)  \ ,
\end{eqnarray}
This exact result can be compared with Eq.~(\ref{eqgue}) in order to find a differential equation for $\widetilde{F}_{N+1}(g^{\prime})$ 
\begin{eqnarray}
 \left( N + (g^{\prime}-1)\frac{d}{dg^{\prime}}\right)\widetilde{F}_{N+1}(g^{\prime}) & =& K_{N}(0,0) \sqrt{\frac{\pi}{g^{\prime}}} \left(\frac{2}{g^{\prime}+1}\right)^{N+1}\nonumber\\ 
 &&\quad \left(\frac{(g^{\prime}-1)^2}{4g^{\prime}}+ N \right) \ .
\end{eqnarray} 
This differential equation can easily be solved. However the solution
\begin{eqnarray}
\widetilde{F}_{N+1}(g^{\prime}) &=& \sqrt{\frac{\pi}{g^{\prime}}}\left(\frac{2}{g^{\prime}+1}\right)^{N}K_{N+1}(0,0)\nonumber\\
     &&\left(1+ \sqrt{g^{\prime}}\left(\frac{g^{\prime}+1}{g^{\prime}-1}\right)^N\int^{g^{\prime}}
 \frac{dx}{\sqrt{x}}\frac{(x-1)^N}{(x+1)^{N+1}}\right)
\end{eqnarray}
is highly complicated. It discourages any attempt to calculate the matrix integral Eq.~(\ref{fmat}) for finite $N$ in the GOE case. To make contact with the results obtained in the main text, we introduce the function 
\begin{equation}
\rho (g^{\prime}) = \sqrt{\frac{2Ng^{\prime}}{(1-g^{\prime})^2}} \ .
\end{equation}
With a change of variables $y= \rho(x)$ in the integral we can write
\begin{eqnarray}
&&\widetilde{F}_{N+1}(g^{\prime}) \ =\ \sqrt{\frac{\pi}{g^{\prime}}}\left(\frac{2}{g^{\prime}+1}\right)^{N}K_{N+1}(0,0)\nonumber\\
 &&\qquad \left(1+\sqrt{\frac{2g^{\prime}}{N}}\left(1+\frac{2 \rho^2(g^{\prime})}{N}\right)^{\frac{N}{2}} \int^{\rho(g^{\prime})} dx 
      \left(1+ \frac{2 x^2}{N}\right)^{-\frac{N}{2}-1}\right) \,
\end{eqnarray}
which coincides with Eq.~(\ref{fmat3}) in the large $N$ limit and for $\beta=2$.

\section{Derivation of Eq.~(\ref{eq1})}
\label{App2}
We define the function 
\begin{equation}
G_N(z) \ \equiv \ \langle \det(H^2+z)^{-\beta/2}|\det H|^\beta \rangle_N  
\end{equation}
where $H$ is a $N\times N$ GOE or GUE random matrix and the brackets denote the corresponding GOE or GUE average. $G_N(z)$ is an analytic function in the cut complex plane ${\mathbb C}\setminus {\mathbb R}_-$. 
In this Appendix we prove the identity
 \begin{eqnarray}
 \label{App21}
G_N(z) 
     & = & L_{N\beta}   
                z^{\beta/2}  \left< \tr\delta(\widetilde{H})\det(\widetilde{H}^2+z)^{-\beta/2} \right>_{N+1}\ , 
\end{eqnarray} 
where $\widetilde{H}$ is a $(N+1)\times (N+1)$ GOE or GUE random matrix. The constant $L_{N\beta}$ is given in Eq.~(\ref{Lcon}). We write the rhs of Eq.~(\ref{App21}) in angle eigenvalue coordinates $\widetilde{H}\to U^{-1}\widetilde{E}U$, where $\widetilde{E}$ is a $N+1\times N+1$ diagonal matrix of the eigenvalues $\widetilde{E}_i$ of $\widetilde{H}$. Since the average is over an invariant function the integral over the diagonalizing group is trivial. The average on the rhs. can now be written as
\begin{eqnarray}
{\rm lhs.}&=& C_{(N+1),\beta}L_{N\beta}   
                z^{\beta/2}  \int d[\widetilde{E}]\sum_{i=1}^{N+1}\delta(\widetilde{E}_i)\prod_{i=1}^{N+1}(\widetilde{E}_i^2+z)^{-\beta/2} \nonumber\\
 &&\qquad\qquad |\Delta_{N+1}(\widetilde{E})|^\beta\exp\left(-\frac{1}{2w^2}\sum_{i=1}^{N+1}\widetilde{E}_i^2\right)\ .
\end{eqnarray}
The power of the Vandermode determinant $\Delta_{N}(x)=\prod_{i<j}(x_i-x_j)$ arises as Jacobian from the coordinate transformation. The constant $C_{(N+1),\beta}$ arising from the group integration can be found in Mehta's book \cite{meh04}. Now the integral over the $\delta$--distribution can be performed.
\begin{eqnarray}
{\rm lhs.}&=& (N+1)C_{(N+1),\beta}\ L_{N\beta}   
                \int d[E]\prod_{i=1}^{N}|E_i|^\beta
                (E_i^2+z)^{-\beta/2} \nonumber\\
 &&\qquad\qquad |\Delta_{N}(E)|^\beta\exp\left(-\frac{1}{2w^2}\sum_{i=1}^{N}E_i^2\right)\ .
\end{eqnarray}
We see that the resulting integral can be written as a GOE (GUE) average over $N\times N$ matrices. This is indicated by using $E_i$ instead of $\widetilde{E}_i$ as integration variables. We go back to Cartesean coordinates $U^{-1}EU\to H$ and find
\begin{eqnarray}
{\rm lhs.}&=& \frac{(N+1)C_{(N+1),\beta}}{C_{(N),\beta}} \ L_{N\beta}   
                \left<|\det H|^\beta
                (H^2+z)^{-\beta/2}\right>_N\ \nonumber\\
           &=&\frac{(N+1)C_{(N+1),\beta}}{C_{N,\beta}} \ L_{N\beta}  G_N(z)\ .
\end{eqnarray}
This is the desired identity with $L_{N\beta}=C_{N,\beta}/(N+1)C_{N+1,\beta}$. Eq.~(\ref{eq1}) is obtained for $z=\epsilon+2iv^2k/\beta$.
  
\section*{References}

\end{document}